\documentclass[times]{speauth}

\usepackage{moreverb}
\usepackage[utf8]{inputenc} %unicode support
\usepackage{listings}
\usepackage{multicol}

\usepackage{graphicx} 
\usepackage{caption} 
\usepackage{subcaption}

\usepackage[colorlinks,bookmarksopen,bookmarksnumbered,citecolor=red,urlcolor=red]{hyperref}

  \definecolor{blue}{rgb}{0,0,0}

\newcommand\BibTeX{{\rmfamily B\kern-.05em \textsc{i\kern-.025em b}\kern-.08em
T\kern-.1667em\lower.7ex\hbox{E}\kern-.125emX}}

\def\volumeyear{2010}

\begin{document}

\title{Analytics-as-a-Service in a Multi-Cloud Environment through Semantically-enabled Hierarchical Data Processing}

\author{Prem Prakash Jayaraman\corrauth\affil{1}, Charith Perera\affil{2}, Dimitrios Georgakopoulos\affil{1}, Schahram Dustdar\affil{3},  Dhavalkumar Thakker\affil{4}, Rajiv Ranjan\affil{5}}

\address{\affilnum{1}RMIT University, Melbourne, 3000 Victoria, Australia.\\
\affilnum{2}Department of Computing,  The Open University, Milton Keynes, MK7 6AA, United Kingdom \\
\affilnum{3}Distributed Systems Group, Vienna University of Technology, Argentinierstrasse 8/184-1, A-1040 Wein, Austria.\\
\affilnum{4}University of Bradford, Bradford BD7 1DP, United kingdom\\
\affilnum{5}School of Computing Science, Newcastle University, NE1 7RU United Kingdom \\
}
\corraddr{RMIT University, Melbourne, 3000 Victoria, Australia.}

\begin{abstract}
A large number of cloud middleware platforms and tools are deployed to support a variety of Internet of Things (IoT) data analytics tasks. It is a common practice that such cloud platforms are only used by its owners to achieve their primary and predefined objectives, where raw and processed data are only consumed by them.  However, allowing third parties to access  processed data to achieve their own objectives significantly increases intergation, cooperation, and can also lead to innovative use of the data. Multi-cloud, privacy-aware environments facilitate such data access, allowing different parties to share processed data to reduce computation resource consumption collectively. However, there are interoperability issues in such environments that involve  heterogeneous data and analytics-as-a-service providers. There is a lack of both -  architectural blueprints that can support such diverse, multi-cloud environments, and corresponding empirical studies that show feasibility of such architectures. In this paper, we have outlined an innovative hierarchical data processing architecture that utilises semantics at all the levels of IoT stack in multi-cloud environments. We demonstrate the feasibility of such architecture by building a system based on this architecture using \textit{OpenIoT} as a middleware, and \textit{Google Cloud} and \textit{Microsoft Azure} as cloud environments. The evaluation shows that the system is scalable and has no significant limitations or overheads. 
\end{abstract}


\maketitle

\section{Introduction}

Recent studies have shown that we generate 2.5 quintillion bytes of data per day \cite{cisco} and this is set to explode to 40 yottabytes by 2020. This will amount to approximately 5,200 gigabytes for every person on earth. Much of these data is and will be generated from the Internet of Things (IoT) \cite{ZMP007}. IoT is a part of the future internet and comprises billions of internet connected objects (ICOs) or \textit{`things'} where each thing can sense, communicate, compute and potentially actuate and can have intelligence, multi-modal interfaces, physical/virtual identities and attributes. ICOs can include wireless/wired sensors, RFIDs, data from social media, smart consumer appliances (TV, smart phone, etc.), smart industries (such as equipments fitted with sensors), scientific instruments (e.g., high energy physics synchrotron) and actuators. The vision of IoT is to allow ‘things’ to be interconnected anytime, anywhere, with anything and anyone, ideally using self-configured paths, networks and services. This vision has led to IoT emerging as a major producer of big data. %Big data applications based on IoT are capable of producing knowledge from billions of real-time data streams required to support timely decision making across numerous application domains.
Today, cloud technologies ~\cite{Wang2:2012, Wang3:2010} provide the ability to store and efficiently process large scale data sets by offering a mix of software and hardware resources with modest operating costs proportional to the actual use (pay-as-you use model) \cite{khalid}.
%Cloud Computing offers an abstracted access to huge pool of resources such as processing, storage and network bandwidth with practically no capital investment and . 
It is well understood that the IoT big data applications need to process and manage streaming data from geographically distributed data sources. The cloud computing model has emerged as a suitable solution to fulfil IoT big data applications' data processing needs.  
%Cloud Computing can offer virtually unrestricted capabilities (e.g. storage and processing) to implement IoT services and application that can exploit the data produced by IoT. 
The cloud essentially acts as a transparent layer between the IoT and applications providing flexibility, scalability and hiding the complexities between the two layers (IoT and applications). The fusion of cloud and IoT into "Cloud of Things" has given rise to the following new cloud computing paradigms (but not limited to): Sensing-as-a-Service, Sensing- and Actuation-as-a-Service, Video-Surveillance-as-a-Service, Big Data Analytics-as-a-Service, Data-as-a-Service, and Sensor-Event-as-a-Service. 
%Given the cloud is a key component in deriving the full benefit of IoT, the term "Cloud of Things" has been coined to describe the amalgamation of Cloud Computing and IoT. 
However, the integrated Cloud of Things approach impose several challenges right from the IoT layer including device discovery, cost-efficient communication, device management and monitoring, interoperability, quality of service and M2M issues to the cloud layer including service discovery and delivery, big data management and analytics, cloud monitoring and orchestration, mobility issues in cloud access, privacy and security and SLA management.  Further, the notion of *-as-a-service model will enable multiple independent operators to provide various services across the CoT layers that will need to be integrated based on application requirements. 
%For example, a provider could own and operate a sensor network infrastructure while another provider provide the analytics components as a service. In such situations, the application will need to integrate the services provided by these providers to satisfy the needs of the end-user.
The prolific rise of IoT and the corresponding ecosystem will soon result in device being owned and operated by independent providers. These solutions will mostly be constrained into independent multiple-cloud provider silos.  A multi-cloud environment consists of several data centres which are geographically and topologically distributed across the Internet \cite{wang2, wang3}.  The focus of this work is to address the challenge of  facilitating multi-cloud data analytics for IoT data originating from things that are owned and operated by multiple service providers. Enabling third parties to access this data and the analytic capabilities can significantly increases the innovation and value of end-user applications.
%This means, the processed data can be securely shared among third parties to achieve their own objectives facilitating a true multi-cloud data analytics paradigm. 
IoT big data applications that need to process and manage streaming data from multiple sources need to exploit the resources hosted across multiple cloud data centres due to  following reasons~\cite{7091808}: 

\begin{itemize}

\item 	\textcolor{blue}{IoT datasets and data sources can be geographically distributed hence moving them to a single centralized data centre could lead to high network communication overhead.}

\item 	\textcolor{blue}{The IoT data storage and processing needs cannot be full-filled by the computational and storage resources offered by any single data centre. For example, in the Azure Cloud, there is a limit of 300 cores per application deployments (i.e. the maximum number of VMs that can be deployed at any instance of time). Clearly, this could lead to serious problems if the IoT datasets flow at a very high volume and velocity.}

\item 	\textcolor{blue}{IoT datasets may be constrained by security and legal policies, i.e., data may not leave a national jurisdiction or can not be streamed into a remote international data centre.}

\end{itemize}

  In this paper, we present hierarchical data analytics model for multi-cloud environments. Our proposed approach allows end-user application to integrate and take advantage of independent infrastructure and analytics service providers. We present a use case to demonstrate the proposed hierarchical and distributed multi-cloud approach to facilitate effective and efficient sharing of analysed data across cloud providers. We use the popular open-source IoT middleware platform namely OpenIoT \cite{soldatos} to demonstrate the feasibility of our approach in multi-cloud environments. Finally, we conduct experimental evaluations on Google Cloud and Microsoft Azure platforms to establish the performance of the proposed hierarchical and distributed multi-cloud approach system.

\textcolor{blue}{It is important to note that our approach is not application dependant. Therefore, it can be generalised in to any application domain where only the analytical functions  employed would need to be differed. Any type of analytical functions can be used on our proposed infrastructure. In this paper, we assume that all the cloud instances who engaged in a given data analytics task are trust-able and verified, before organise them into a certain hierarchical composition in order to support a given application.}

%To demonstrate the feasibility of our approach, we perform experimental evaluations using a popular IoT middleware called OpenIoT \cite{soldatos}.}

\section{Motivation: Analytics-as-a-Service}

In sensing-as-a-service \cite{ZMJ002} model,  data is exchanged seamlessly among data producers (owners) and consumers via the cloud resources. Data producers are owners of the IoT devices (products) and deploy them in their environments. These IoT products sense, analyse and perform actuation to solve the needs of the data owners. While this data normally resides in individual silos, sensing-as-a-service model promotes the sharing of data (liberating data from silos) allowing data consumers to access the data using secure mechanisms. For example, a plant biologist studying the spread of certain diseases in plants may want to know the list of affected farms to better understand the trajectory of the diseases. In this case, the aim of the biologist is not to identify individual farms, but a while set of farms in specific areas. When the number of data providers and consumers increase, there is a need to develop an open data market. The data from this market may not necessarily freely available \cite{6803135} (may follow the cloud computing pay-as-you-go model) but the metadata description the data would be. The meta data will enable users and other services to discover relevant data stored in data owner silos.

Analytics-as-as-Service refers to next generation IoT data processing applications where third party will be responsible for hosting IoT Analytics and data processing applications (e.g., detecting events from video camera feeds, detecting events from smart home sensors, etc.) on private/public cloud infrastructures. These analytics applications will be offered to end-users under pay-as-you-go-model. Currently, such a service model is offered for cloud-based hardware (CPU, Storage, and Network) and software (Databases, message queuing systems, etc.) resources by providers such as Amazon Web Services. Providers such as SalesForce.com offers pay-as-you-go model for ERP and CRM applications. However, ERP and CRM applications are fundamentally different from IoT Analytics applications.
%The analytics-as-a-service model arises from the sensing-as-a-service model promoting the creation of analysed open data markets (in multi-cloud environments). For example, an expert in plant biology (biologist) may have the knowledge to determine (analytics) plant growth by interpreting data obtained from sensorised fields. Providing this analytics as a service will be invaluable to farmers (data producers/owners). 
Moreover analytics-as-a-service model introduces further complexities as there is need to describe not only the data but also the analytics performed on the data. Further, when data analytics exists as data silos within independent data owner clouds, there is a need to develop systems that can function across multiple cloud providers. Such systems will inherently require the following capabilities namely 1) ability to interoperate via standard interfaces 2) ability to describe data 3) support for machine to machine communication and 4) ability to describe the analytics built on the acquired data.

Another advantage provided by analytics-as-a-service model is that it supports knowledge sharing while reducing the privacy risks. Due to the fact that this model does not share raw data, it eliminates the risks associates with sharing raw data such as anonymised sharing of analysed data, enforce restrictions on data storage location etc. Another advantage is the savings of computational resources due to the elimination of redundant data processing. This means that when one cloud IoT platform perform a certain data processing task over data, the recipient cloud platforms does not required to perform the same data processing task again. For example, one IoT cloud platform may collect data form motion sensors and cameras to determine how much time in average a person may wait in a certain queue. One such data processing is done, the recipient cloud can take average waiting time as an input. We elaborate on this example in Section \ref{sec:Case_Study} when we present the use-case scenario. Further, analytics-as-a-service model also reduces the data communication requirements. Typically, raw data is large in term of size. However, the processed data is significantly smaller that raw data. Therefore, the amount of data that need to be transferred from one cloud to another reduces drastically by saving network communication bandwidth and costs.

\section{Current State of the Art: Processing Distributed Internet of Things Data }
Existing big data processing technologies and data centre infrastructures~\cite{Wang1:2013} have varied capabilities with respect to meeting the distributed IoT data processing challenges. In this section we summarize capabilities of existing technologies based on the review given in our past work~\cite{7091808}.  The proposed analytics-as-a-service model is expected to be extensively leverage these technologies. We have reviewed literature under six different themes: 1) basic data centre cloud computing infrastructure service stacks, 2) massive data processing models and frameworks, 3) trusted and integrated data management services across data centres, 4) data-intensive workflow computing, 5) benchmarking, application kernels, standards and recommendations, and 6) sensing middleware in the Cloud.

\textbf{1) Basic data centre cloud computing infrastructure service stacks}

Commercial or public data centres, for example,  Amazon Web Services and Microsoft Azure offer computing, storage, and software resources as remotely programmable  cloud services via Application Programming Interface (API). These resources are orchestrated by deploying virtualization software/middleware stacks. It is well understood that virtualization allows data centre providers to get more out of physical resources by allowing multiple instances of virtual cloud resources to run concurrently.  For example, virtual machine orchestration systems such as Eucalyptus and Amazon EC2; image management tools such as FutureGrid image repository \cite{FutureGrid}; massive data storage/file system such as GFS, HDFS, and Amazon S3; and data-intensive execution framework including Amazon Elastic Map Reduce. In addition, FutureGrid\footnote{http://FutureGrid.org/} and OpenStack also provide software stack definition for cloud data centres.

On the other hand, private data centres are constructed  typically by combining multiple types of software tools \& services. These software can include, cluster management systems such as Torque, OSCAR, VMWare's vCloud and/or vSphere suites and SLURM (Simple Linux Utility for Resource Management); parallel file/storage systems such as SAN/NAS\footnote{ http://capitalhead.com/articles/san-vs-das-a-cost-analysis-of-storage-in-the-enterprise.aspx}, Lustre\footnotetext{http://wiki.lustre.org/}; as well as data management systems such as BeSTMan\footnote{https://sdm.lbl.gov/bestman/} and dCache\footnote{http://www.dcache.org/}. Apart from, some private data centres are enabled for resource sharing with Grid computing middleware, such as Globus Toolkits, Unicore, and gLite. In general access to private data centre resources is restricted to known group of application administrators and users due to stringent security and privacy concerns.

\textbf{2)	Big data processing models and frameworks}

Big Data Processing Frameworks include software frameworks that enable creation of big data application architecture \cite{wang4}. These frameworks can be classified as follows: 
\begin{itemize}
\item	Large-Scale Data Mining frameworks (FlexGP, Apache Mahout, MLBase, Yahoo SAMOA) implement a wide range of Data Mining (DM) algorithms (clustering, decision trees, latent Dirichlet allocation, regression, Bayesian) to analyse massive data sets (historical and streaming) in parallel, by exploiting distributed resources.
\item	Distributed Message Queuing frameworks (Amazon Kinesis, Apache Kafka) provide a reliable, high-throughput, and low-latency system of queuing real-time streams of data.
\item Parallel and Distributed Data Programming frameworks (Apache Hadoop, Apache Storm). Such frameworks enable development of distributed applications that deal with large sets of cloud resources to parallel process massive amounts of historical and streaming data \cite{wang1:2015, wang4}. The large scale DM frameworks mentioned above are generally implemented on top of parallel and distributed data programming frameworks. Low-level distributed system management complexities (task scheduling, data staging, fault management, inter-process communication, result collection) are automatically taken care of by these frameworks.
\item Data Store frameworks are categorised as NoSQL and SQL. NoSQL frameworks (MongoDB, HyperTable, Cassandra, Amazon Dynamo) support access based on transactional programming primitives, where an exact key allows search for an exact value. Such predetermined access patterns lead to better scalability and predictions of performance, which is suitable for storing large amounts of unstructured data (e.g. social media postings). SQL data stores (MySQL, SQL Server, PostGreSQL) manage data in relational tables, where the generic Structured Query Language can be used to manipulate data (insert, delete, update). In essence, SQL Data Stores are more effective than NoSQL stores, where transactional integrity (ACID properties) is a strict requirement. Future big data applications are likely to use both NoSQL and SQL data stores, driven by data varieties and querying needs. SQL Engines (Apache Hive, Apache Pig) enable the querying of data across a variety of cloud storage resources including Amazon S3 and Hadoop Distributed File System (HDFS) based on structured query language.
\end{itemize}

\textbf{4)	Data-intensive workflow orchestration framework}

Typical workflow frameworks for managing scientific big data applications includes Pegasus, Kepler, Taverna, Triana, Swift, and Trident. Trdationally, in service computing domain  orchestration with BPEL and YAWL \cite{quteprints80697} has been extensively explored. On the other hand, service choreography has been done using WS-CDL\footnote{http://www.w3.org/TR/ws-cdl-10/}. More recently, orchestration frameworks such as YARN (Yet Another Resource Negotiator~\cite{Vavilapalli:2013}) and Mesos~\cite{Hindman:2011} have emerged for coordinating IoT data analytics workflow tasks across multiple big data processing frameworks (e.g. Apache Hadoop, Apache Storm, etc.). 

\textbf{5)	Benchmark, application kernels, standards and recommendations}

Several benchmarks and application kernels have been developed, for example, Graph 500 (graph500.org/), Hadoop Sort\footnote{http://wiki.apache.org/hadoop/Sort}and Sort benchmark (sortbenchmark.org), MalStone \cite{Bennett}, Yahoo! Cloud Serving Benchmark\footnote{http://research.yahoo.com/Web\_Information\_Management/YCSB}, Google cluster workload\footnote{http://code.google.com/p/googleclusterdata/}, TPC-H benchmarks (www.tpc.org/tpch), BigDataBench, BigBench, Hibench, PigMix, CloudSuite, and GridMix powered by the needs of analyzing the performance of different big data workloads. These benchmark suites model workloads for stress testing one or more categories of big data processing frameworks such as Apache Hadoop and Apache Mahout. In the current generation of  framework suites, BigDataBench and BigBench are the most comprehensive ones. This is due to the fact that they incorporate big data workload models for variety of processing frameworks including NoSQL, DBMS, SPEs and batch processing frameworks. Mainly, BigDataBench targets the application domains such as search engine, social network, and e-commerce. Having said that, their is limited benchmarks and application kernels available for heterogeneous data centers and IoT data tyoes. Specially, there is no consensus on available performance benchmarking for executing large-scale IoT applications across distributed data centers. 	Literally, the absence of inter-centre benchmark and standards need to be the primary research agenda for the future. As of now, international organizations include NIST, OGF, DMTF Cloud working group, Cloud Security Alliance, and Cloud Standards Customer Councilare all working on cloud standards (occi-wg.org/)\footnote{http://www.dmtf.org/standards/ovf}.

\textbf{6)	Sensing Middleware in the Cloud}

Over the last few years, number of IoT cloud has been made their way in the sensing middleware marketplace. Thingworx (thingworx.com) and Xively (xively.com) are cloud-based online platforms that process, analyse, and manage sensor data retrieved through a variety of different protocols. HomeOS \cite{Z1058} is a platform that supports home automation. HomeOS is a software platform which can be installed on a normal PC. As with the smartthings platform, applications can be installed to support different context-aware functionalities (e.g. capturing an image from a door camera and sending it to the user when someone rings the doorbell). Lab-of-things \cite{Z1059} is a platform built for experimental research. It allows the user to easily connect hardware sensors to the software platform and enables the collection of data and the sharing of
data, codes, and participants. However, most of these platforms hosted on the cloud by their owners and customers have no choice on the cloud technologies used. There are a few open source IoT platform developed by both research community (e.g. OpenIoT \cite{soldatos}) and industrial players (e.g. WSO2 IoT-wso2.com/landing/internet-of-things/) that can be hosted any cloud available in the market today. Therefore, in this paper, we used OpenIoT as the IoT platform of choice to develop the prototypes.

\section{Hierarchical Data Analytics in Multi-Clouds}
In this section, first, we explain what hierarchical data analysis means in multi-cloud environment and its important feature and characteristics. We then present the widely used open-source IoT platform OpenIoT and describe its features that enable multi-cloud hierarchical processing. The presented OpenIoT platform is driven by semantic web concepts and hence incorporates extensive use of ontologies to define devices and services. This feature of OpenIoT, which will be presented in detail is the foundation for achieving the hierarchical multi-cloud data analytics model.

\begin{figure}[t!]
 \centering
% \vspace{-0.23cm}
 \includegraphics[scale=0.90]{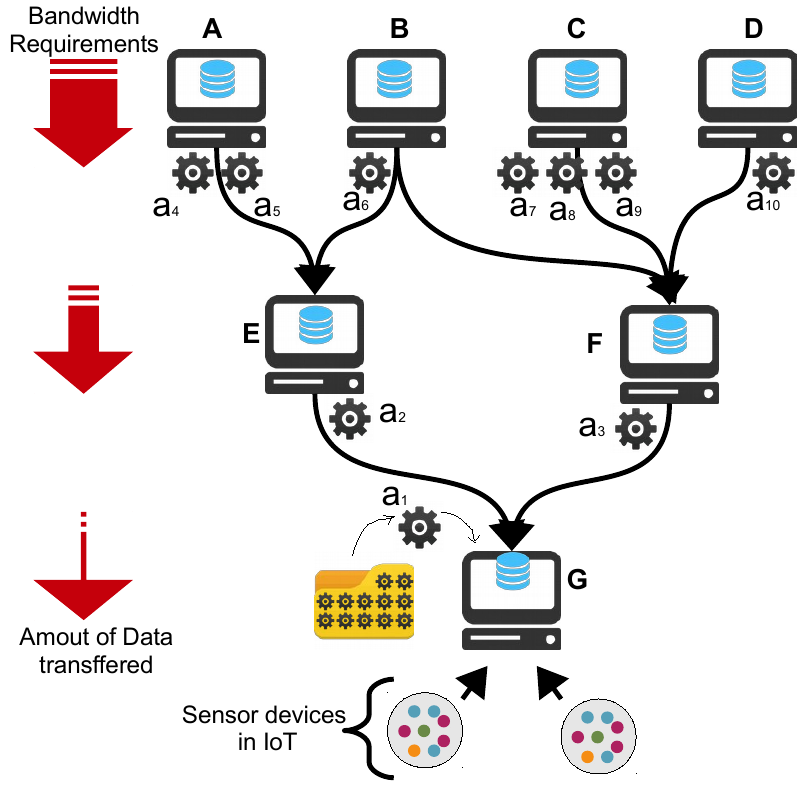}
%\vspace{-0.43cm}	
 \caption{Theoretical View of Hierarchical \\Data Analytics }
 \label{Figure:Theory}	
%\vspace{-0.53cm}	
\end{figure}
%\add{Charith: Please add a discovery component to the figure 1. To show that each of the service/data provider will register with a discovery component}

Let us consider the Figure \ref{Figure:Theory}. It is important to note that hierarchical data analytics does not means that communication network has to be hierarchical. Hierarchical data  analysis can happen in any type of network. The fundamental idea is as follows. First, data is captured by leaf nodes. In Figure \ref{Figure:Theory}, nodes A, B, C, and D can be considered as leaf notes which are responsible for gathering data streams generated by different sources. Data sources could be hardware sensors (e.g. temperature sensor) or a virtual sensors (e.g. calling a weather service). First, the leaf nodes may analyse the data they gathered. Each node may have their own data analytical capabilities (as denoted in a$_{1}$...a$_{10}$) based on the library of data analytics tools they have access to. Once data analytics are applied by leaf node, the data is transferred to the next layer of nodes (i.e. node E and F). These nodes will run another set of analytics over the incoming data streams and generate more abstract outputs (i.e. a data stream). Finally, E and F nodes transfer their outputs to node G.

It is important to note that data processing does not follow any particular layered structure. The idea is to perform analytics  in a node and pass the results onto another node to perform another set of analytics. As a result A, B, C, D does no have to be in the same layer. One stream of data may directly be sent to node A without sending them to node E if the analytics performed in node E is not required by the node A.

In both sensing-as-a-service model and analytics-as-a-service models, nodes are collecting  and processing data  in order to achieve  their own objective. Hierarchical data analytics in multi-cloud environment occurs, when a given node  does not have access to required data (e.g. node G). In such occasions, initiation node sends requests to other nodes in order to get access to the data it requires. Further, as shown by red arrows in Figure \ref{Figure:Theory}, the amount of data need to be transferred between nodes as well as the bandwidth requirement get reduced at each layer. Primarily the reason for this is that each layer performs some-kind of analytics over the data and generates  more aggregated results. For example, an average function may aggregate data over 5 minutes and generate a single tuple. In another instance, a function may combine sensor data from video cameras to identify the number of people entering into a certain area over an hour. Without sensing streaming video feeds, each processing node may only stream the number count to the next node in the hierarchy. The proposed model has several advantages namely: 
\begin{itemize}
\item It facilitates integration of services across various layers
\item It allows seamless integration of data producers and consumers staying agnostic to infrastructure and technologies
\item It is a platform to build complex end-user applications without owning the data production infrastructure nor the data processing tools/infrastructure
\item Allow seamless discovery of service provider capabilities that can be implemented using many mechanisms including semantic discovery, probabilistic discovery, SOA-style discovery etc.
\end{itemize}

\subsection{OpenIoT: An Open source middleware for Internet of Things}

The OpenIoT middleware \cite{soldatos} is a versatile blueprint architecture for collecting and processing data from Internet of Things data sources. OpenIoT provides an innovative complete IoT stack platform for IoT/cloud convergence which enables: (A) The integration and streaming of IoT data and applications within cloud computing infrastructures; (B) The deployment of semantically interoperable applications in the cloud; (C) The implementation of mainstream cloud computing concepts and properties in the IoT domain, including the concept of <Sensing-as-a-Service> (i.e. on-demand, utility-based access to IoT services) and the concept of pay-as-you-go for IoT applications; (D) Handling of mobile sensors (e.g., smart phones) and associated QoS parameters (e.g., energy efficiency). OpenIoT currently uses standard communication protocols such as TCP/IP and RESTful architecture to enable communication between the different components. However, it is an open framework with support for any new protocols such as CoAP.

\subsubsection{OpenIoT: Architectural Overview}
\label{openiot-arch}

The OpenIoT architecture is comprised of seven main elements that belong to three different logical planes, as illustrated in Figure \ref{Figure:OpenIoTArch}. These planes are the Utility/Application Plane, the Virtualized Plane and the Physical Plane which include the following modules:

\begin{figure}[b!]
	\centering
	% \vspace{-0.23cm}
	\includegraphics[scale=0.25]{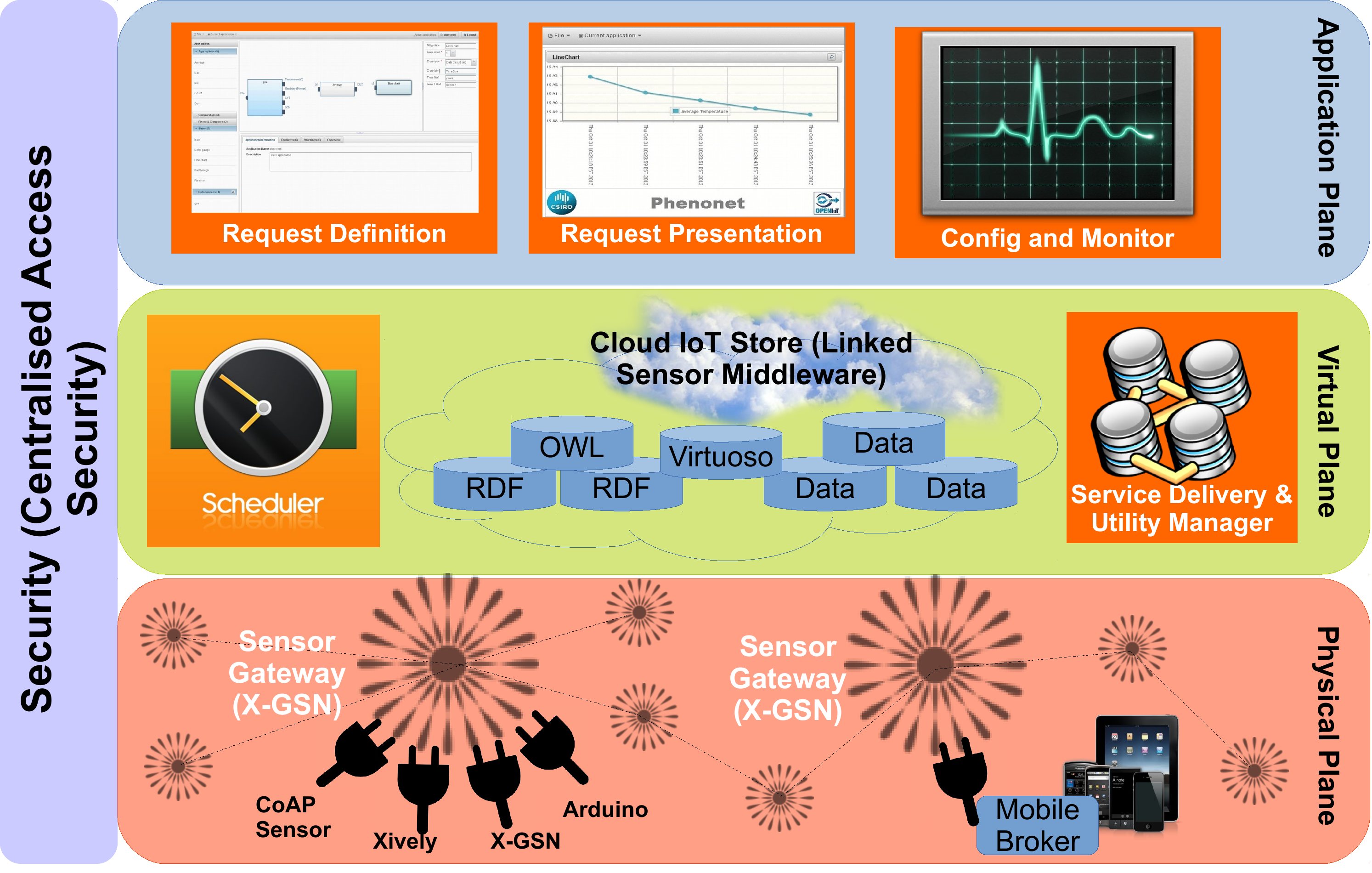}
	%\vspace{-0.43cm}	
	\caption{OpenIoT Architectural Overview}
	\label{Figure:OpenIoTArch}	
	%\vspace{-0.53cm}	
\end{figure}

	\textbf{Utility/Application Plane:} The utility and application plane is responsible for managing interaction with end-user applications. In particular, it provides a set of tools and interfaces that users can use to deploy IoT application on-the-fly.  It comprises the following key components namely:
	\begin{itemize}
		\item The Request Definition enables the specification of service requests to the OpenIoT platform. It comprises a set of services for specifying and formulating such requests, while also submitting them to the Global Scheduler. This component can be realised using a feature rich GUI (Graphical User Interface) allowing user interaction or via APIs for machine to machine communication.
		\item The Request Presentation is responsible for visualising  the outputs of an IoT service. This component creates  mashups from the service decribption in order to facilitate presentation of analysed data.
		\item The Configuration and Monitoring component enables the management and configuration of functionalities over the sensors and the (OpenIoT) services that are deployed within the OpenIoT platform. Moreover, it enables the user to monitor the health of the different deployed modules.
	\end{itemize}

	\textbf{Virtualized Plane:} The virtual plane is responsbile to bridge the device layer (physical) to the application layer. The virtual plane in most cases is deployed on cloud environments and is responsible for providing core functionalities and services to the physical and application layer. Note that the cloud infrastructure could be either a public infrastructure (such as the Amazon Elastic Compute Cloud (EC2)) or a private infrastructure (e.g., a private cloud deployed based on Open Stack (http://www.openstack.org/)). It comprises the following components
	
	\begin{itemize}

		\item The Directory Service (LSM-Light), keeps information about all the sensors and services that are available in the OpenIoT platform.  It also provides the means (i.e. services) for registering sensors and services with the directory, as well as for the look-up (i.e. discovery) of sensors and services. The architecture specifies the use of semantically annotated descriptions of sensors as part of its directory service. This component is developed by extending the W3C SSN ontology \cite{soldatos} allowing representation of both sensors and their corresponding services respectively. The directory service can be characterized as a sensor cloud, given that it primarily supports storage and management of sensor data streams (and of their metadata). This component of OpenIoT is vital to the relational of the proposed hierarchical multi-cloud data analytics approach and will be discussed in detail in the following section.
		
		%The OpenIoT open source implementation is based on an enhanced version of the W3C SSN ontology, which is integrated as part of the LSM (Linked Sensor Middleware). As a result of this implementation technology, semantic Web techniques (such as SPARQL and RDF (Resource Description Format)) and ontology management systems (e.g., Virtuoso) are used for querying the directory service. Furthermore, the exploitation of semantically annotated sensors enables the integration of data streams within the Linked Data Cloud, thereby empowering Linked Sensor Data. 
		\item The Global Scheduler, processes all the requests for on-demand deployment of services and ensures their proper access to the resources (e.g. data streams). This component undertakes the task of parsing the service request and accordingly discovering the sensors that can contribute to its fulfilment. It also selects the resources, i.e., sensors that will support the service deployment, while also performing the relevant reservations of resources.

		\item The Service Delivery \& Utility Manager (SDUM), which performs a dual role. On one hand, it combines the data streams as indicated by service workflow description, in order to deliver the requested service. To this end, this component makes use of the service description and the resources identified and reserved by the (Global) Scheduler component. On the other hand, this component acts as a service metering facility, which keeps track of utility metrics for each individual service. This allows utility-based metering to facilitate the development of application using service provided by disparate providers.
	\end{itemize}

	\textbf{Physical Plane:} The physical plane refers to the devices deployed in the physical environment. This can include real hardware sensors and virtual sensors. This layer is responsible for managing interactions between the device layer and the upper layers (virtual and application). This layer enables both sensing and actuation capabilities. This layer comprises the following component
	\begin{itemize}
		\item The Sensor Middleware (Gateway), which collects, filters and combines data streams stemming from virtual sensors (e.g. signal processing algorithms, information fusion algorithms and social media data streams) or physical sensing devices (such as temperature sensors, humidity sensors and weather stations). This middleware acts as a hub between the OpenIoT platform and the physical world, since it enables access to information stemming from the real world. Furthermore, it facilitates the interfacing to a variety of physical and virtual sensors such as IETF COAP compliant sensors (i.e. sensors providing RESTful interfaces), data streams from other IoT platforms (such as https://xively.com) and social networks (such as Twitter).  Among the main characteristics of the sensor middleware is its ability to stream W3 SSN compliant sensor data in the cloud. The Sensor Middleware is deployed on the basis of one or more distributed instances (nodes), which may belong to different administrative entities. The prototype implementation of the OpenIoT platform uses an enhanced/extended version of the GSN middleware (namely X-GSN, which is currently as a module of the OpenIoT open source project). However, other sensor middleware platforms could be also used in alternative implementations and deployments of the OpenIoT architecture.
	\end{itemize}
	
	\textbf{Security Plane}: The security plane cuts across the OpenIoT architecture stack ensuring an end-to-end security mechanism. The platform uses a token-based authentication system supported by role-based access control for authentication, authorisation and identity management.
						\begin{figure*}[t!]
							\centering
							% \vspace{-0.23cm}
							\includegraphics[scale=0.5]{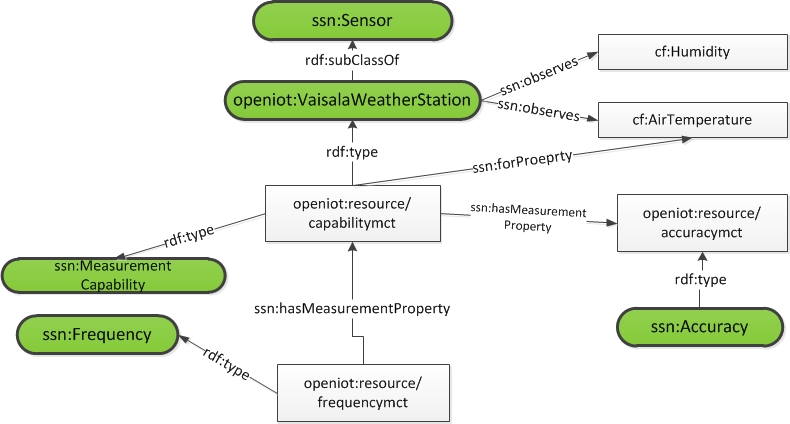}
							%\vspace{-0.43cm}	
							\caption{Sensor Description based on SSN}
							\label{Figure:RDF}	
							%\vspace{-0.53cm}	
						\end{figure*}

\subsection{Hierarchical Multi-Cloud Data Analytics using OpenIoT}
The OpenIoT system is driven by semantic web technologies. It extensively uses an enhanced version of the W3C SSN ontology namely OpenIoT ontology \cite{Soldatos2015} to for semantics annotation of data at each layer of the IoT stack i.e. device layer, virtual layer and the application layers. OpenIoT exploits other semantic web technologies such as Linked Data\cite{bizer2009linked} for dynamically linking related sensor data sets with corresponding services and vice-versa and Resource Description Framework (RDF), Web Ontology Language (OWL) and Simple Protocol and RDF Query Language (SPARQL) for for semantic modelling, representation, storage and retrieval of sensors and services. In this section, we will present the features of the OpenIoT architecture that enables the realisation of multi-cloud data analytics applications.

The virtual layer services namely LSM-Light, Scheduler and SDUM are at the heart of the OpenIoT architecture that enables the following capabilities namely: 1) Ability to register sensors with semantic descriptions, 2) Ability to  register service that are composed by the user/application and 3) a discovery service that enables semantic discovery of sensors and service. A \textit{service} in OpenIoT is defined as a specification that defines the set of analytical operation to be performed on a stream of  sensor data and the respective visual presentation.

\textbf{Description of Devices:}
The OpenIoT Ontology extends the W3C SSN ontology enabling it to describe and register devices (sensors and things) with the virtual layer. Figure \ref{Figure:RDF} presents an example of a partial sensor description. The RDF below describes a sensor namely a \textit{Vaisala Weather Station} that has the capability to measure temperature and humidity.

\textbf{Description of Services:}
The OpenIoT Service Description specification (OSDSpec) is capable of describing in detail the service composed by the user/application. The OSDSpec is modelled in the OpenIoT ontology and is stored/managed by the directory service and scheduler components of the virtual layer. This OSDSpec allows the service to be described in detail including query control features such as query schedule, permissions on the query etc. Listing 1 is an example of an OpenIoT OSDSpec.

\textbf{Discovery and Invocation of Devices and Services}

Once the devices and services are registered with the virtual plane namely the directory service, the directory service along with the scheduler and SDUM are used to discover and invoke composed services. Listing 2 presents a sample SPARQL query that is used to perform semantic discovery for devices (things) within a given location. The query also takes additional parameters such as \textit{SensorType}, \textit{SensorClass} to perform more efficient discovery. The discovery service is also used to discover services e.g. an analytic service offered by a service provider. Together, the virtual planes enables application to discover services offered by independent sensor infrastructure owners and analytics service providers.

\lstset{language=XML,caption={Sample OpenIoT Service Specification},label=OSDSpec}

\begin{lstlisting}[frame=single, breaklines=true, float=*][t!] 
	<?xml version="1.0" encoding="UTF-8"?>
	<osd:OSDSpec xmlns:st="http://www.w3.org/2007/SPARQL/protocol-types#"
	xmlns:vbr="http://www.w3.org/2007/SPARQL/results#"
	xmlns:rdf="http://www.w3.org/1999/02/22-rdf-syntax-ns#"
	xmlns:osd="http://www.openiot.eu/osdspec"
	xmlns:xsi="http://www.w3.org/2001/XMLSchema-instance">
		<osd:OAMO name="name0">
			<osd:OSMO name="name1">
				<osd:queryControls>
				<osd:QuerySchedule>
			</osd:QuerySchedule>
			<osd:reportIfEmpty>false</osd:reportIfEmpty>
			</osd:queryControls>
			<osd:requestPresentation>
				<osd:widget widgetID="http://www.oxygenxml.com/">
					<osd:presentationAttr name="name2" value="value0"/>
					<osd:presentationAttr name="name3" value="value1"/>
				</osd:widget>
				<osd:widget widgetID="http://www.oxygenxml.com/">
					<osd:presentationAttr name="name4" value="value2"/>
					<osd:presentationAttr name="name5" value="value3"/>
				</osd:widget>
			</osd:requestPresentation>
			<st:query-request>
				<query>query0</query>
			</st:query-request>
			<st:query-request>
				<query>query1</query>
			</st:query-request>
			</osd:OSMO>
		</osd:OAMO>
	</osd:OSDSpec>
\end{lstlisting}

\lstset{language=XML,caption={Sample Device Discovery Query},label=DeviceDiscoveryQuery}
\begin{lstlisting} [frame=single, breaklines=true, float = *] 
SELECT ?graphNode_2197552479500_sensorId
FROM <http://openiot.eu/OpenIoT/sensormeta#>
WHERE
{
?graphNode_2197552479500_sensorId <http://www.w3.org/1999/02/22-rdf-syntax-ns#type> <http://demo.org/ns#TestType> .
<http://demo.org/ns#TestType> <http://www.w3.org/2000/01/rdf-schema#subClassOf> <http://purl.oclc.org/NET/ssnx/ssn#Sensor> .
?graphNode_2197552479500_sensorId <http://www.loa-cnr.it/ontologies/DUL.owl#hasLocation> ?graphNode_2197552479500_loc .
?graphNode_2197552479500_loc geo:geometry ?graphNode_2197552479500_geo .
?graphNode_2197552479500_loc geo:lat ?graphNode_2197552479500_lat .
?graphNode_2197552479500_loc geo:long ?graphNode_2197552479500_lon .
FILTER (<bif:st_intersects>(?graphNode_2197552479500_geo, <bif:st_point>( 6.635227203369141, 46.52119378179781), 15)) .
}
\end{lstlisting}

The virtual plane components also provide API interfaces to invoke the discovered services. The key contribution of the proposed multi-cloud model is to promote interoperability among different data and analytic service providers. This is achieved by the discovery service combined with the API allowing the development of the multi-cloud data analytics applications.

\section{Experimentations and Evaluations}
\label{sec:Case_Study}
In this section, we present a real-world usecase scenario where we demonstrate the importance of hierarchical data processing in multi-cloud environments. Then, we describe the experimental test-bed implemented using the OpenIoT system in order to validate the feasibility and conduct performance evaluations.

\subsection {A Case Study}

\begin{figure}[t!]
 \centering
% \vspace{-0.23cm}
 \includegraphics[scale=0.54]{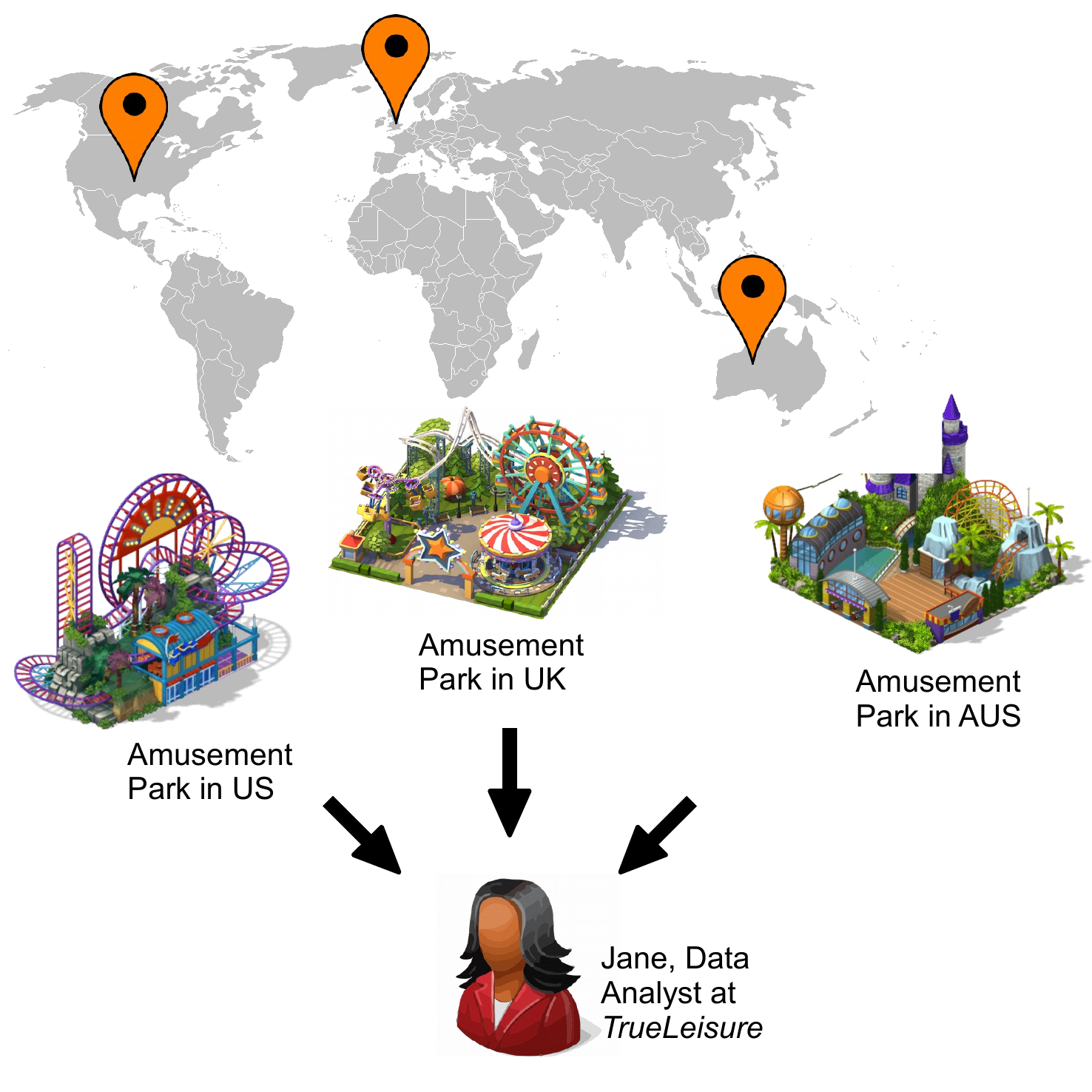}
%\vspace{-0.43cm}	
 \caption{A Case Study: Service Quality Monitoring \\of Amusement Park Chain}
 \label{Figure:Usecase}	
%\vspace{-0.53cm}	
\end{figure}

\textit{TrueLeisure} is company that operates different types of entertainment attractions. Among them they have franchised their amusement park chain. As depicted in Figure \ref{Figure:Usecase}, currently Amusement parks are located in United States, United Kingdom, and Australia. These amusement pars are fully owned and operated by the franchisees. However, \textit{TrueLeisure} continuously monitor and assess the service qualities and several other aspects of each of the amusements part. \textit{TrueLeisure} takes these assessment seriously as their brand image is dependent on the quality of the services provided by the franchisees.

Jane is a data analyst overseeing the quality assessment tasks of amusement parks at \textit{TrueLeisure}. She is responsible for continuously monitoring the service quality parameters. In addition to Jane, each of the franchisees also have their own data analysis and quality control division where they also monitor their own quality parameters. All the amusements parks are augmented with a large number of sensors that collects various types information such as environmental parameters (e.g. temperature, humidity, pressure), crowd movements, usage and demand of each rides and attractions, operational status of machinery used in the amusement part, etc. Each of the amusement parks have deployed their own IoT platforms to which sensors are connected. Conceptually, a query would look like SELECT  AVG(WaitingTime) FROM United States, United Kingdom, Australia. The importance of this type of abstraction is that  Jane does not need to know how to find waiting times in each location where each location may employ different technological means to acquire different types of sensors data to derive waiting times.

\begin{figure}[b!]
	\centering
	% \vspace{-0.23cm}
	\includegraphics[scale=0.54]{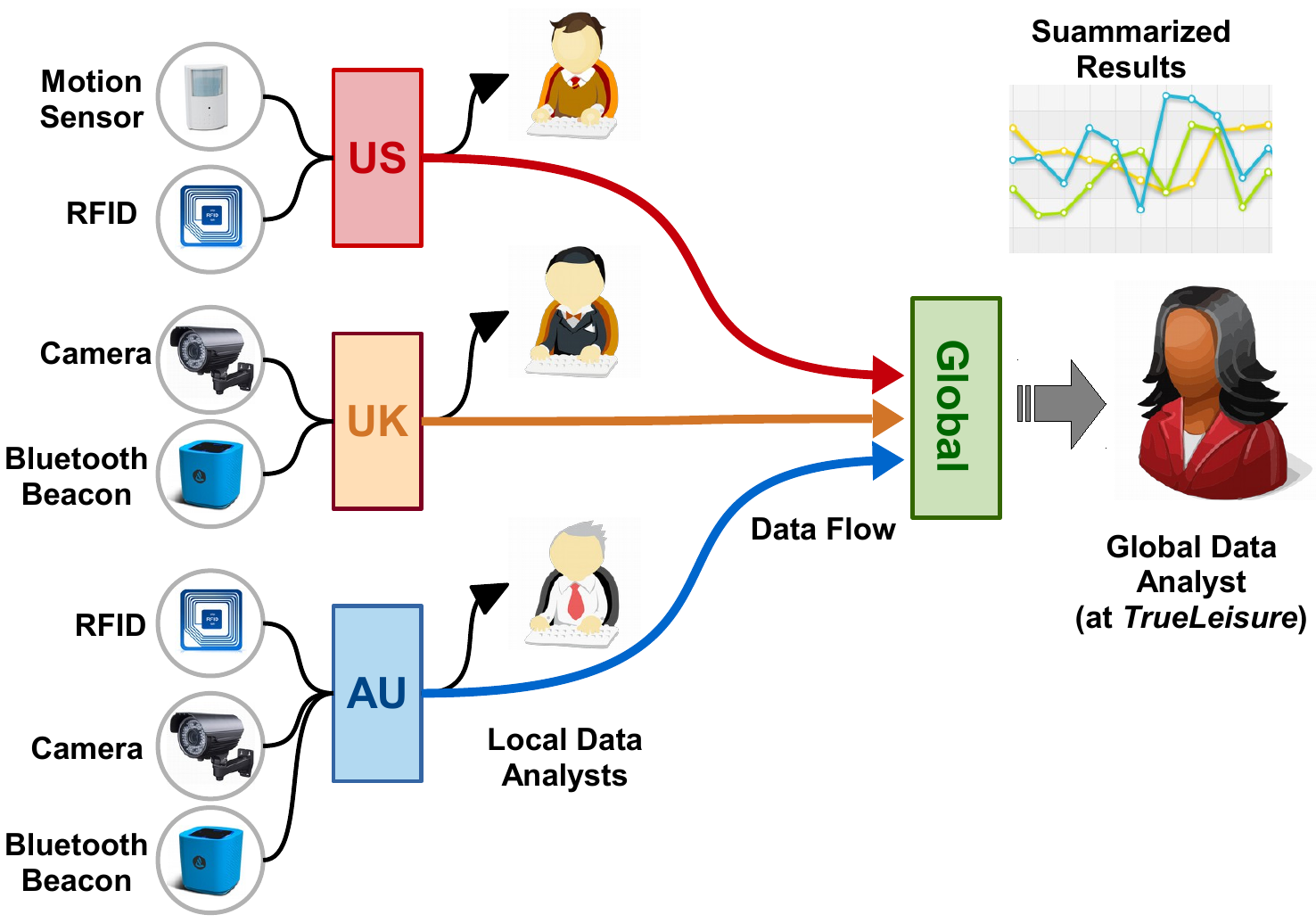}
	%\vspace{-0.43cm}	
	\caption{Data Flow in Hierarchical Data \\Processing}
	\label{Figure:DataFlow}	
	%\vspace{-0.53cm}	
\end{figure}

\begin{table*}[!t]
	\centering
    \renewcommand{\arraystretch}{1.4}
    \normalsize
	\begin{tabular}{ l | c | r }
	    
		\hline
		ServerName & Location/Zone & Configuration \\
		\hline
		OpenIoT-1-Azure & 	Australia East & Standard Instance, A3(4 Cores, 7GB Memory) \\
		OpenIoT-2-Azure & 	Australia East & Standard Instance, A2(2 Cores, 3.5GB Memory) \\
		OpenIoT-1-Google & asia-east1-a & n1-standard-2 (2 vCPUs, 7.5 GB memory)
		
	\end{tabular}
	\caption{OpenIoT Implementation Details}
	\label{cloud-spec:table}
\end{table*}

One of the important service quality parameter is \textit{`waiting time'}. This is a main contribution factor towards customer satisfaction. Local quality assessment team continuously measures the crowd waiting time of each ride and attraction within their own amusement park. The raw  data generated by sensors such as motion sensors, cameras, Bluetooth beacons, RFID tags are used to calculate these waiting times. By measuring waiting times, local data analysis team can recommend their operational division about any bottleneck within the park so the management can take necessary actions to eliminate those to increase customer satisfaction. From Jane's perspective, who is responsible for overseeing entire portfolio of amusement parks at \textit{TrueLeisure}, she is only interested in the \textit{big picture}. That means Jane would like to create a single parameter  of waiting time (i.e. overall waiting time) by combining individual waiting times (i.e. individual waiting time for each ride or attraction) together. As a results, she will have three measures where each represent waiting time of each amusement park locates in United State, United Kingdom and Australia. By plotting these measures in a line chart , Jane can view how waiting time varies in real-time. Jane will report these high-level measures to her corporate management so \textit{TrueLeisure} can discuss with their franchises on future development of their theme parks efficiently and effectively. Figure \ref{Figure:DataFlow} illustrate how data is being collected, processed and transferred in such a scenario using the proposed hierarchical data analysis in a multi-cloud environment. This scenario is a typical example of data producers, analysis service providers and data consumers operating and managing their own infrastructure (each theme park) and applications integrating these services to address specific requirements (Jane interested in overall performance of each theme park). 

\subsection{Experimental Setup}

\begin{figure}[b!]
	\centering
	% \vspace{-0.23cm}
	\includegraphics[scale=1.0]{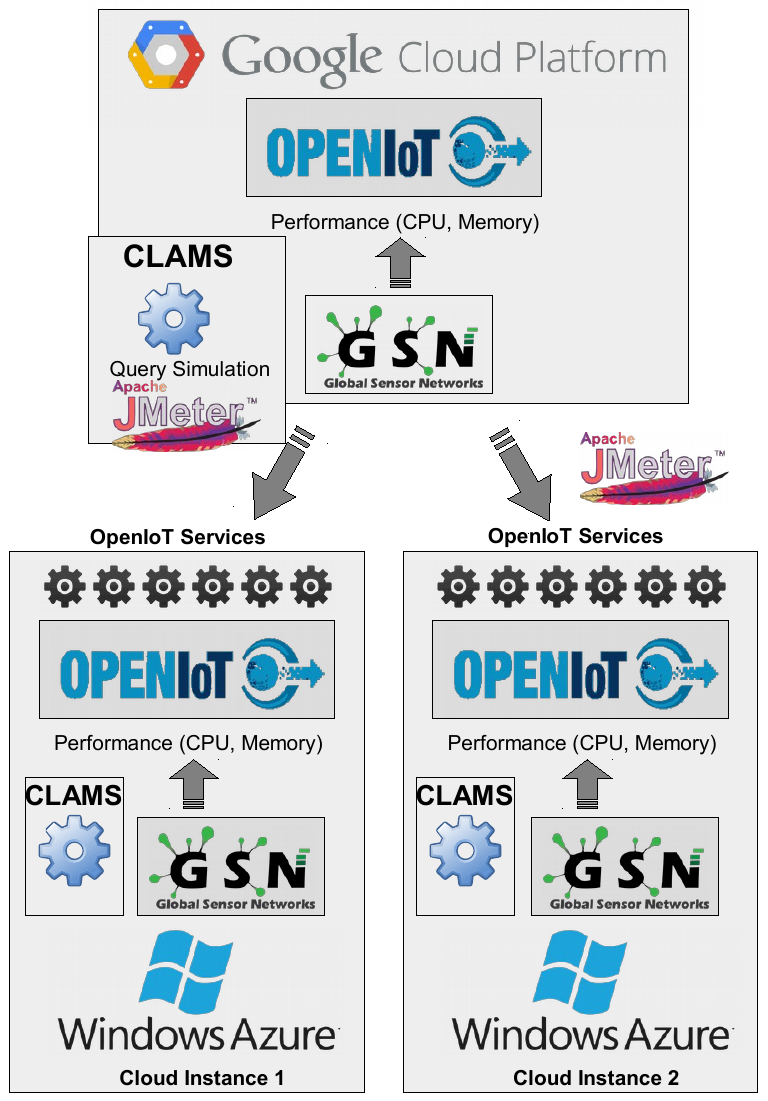}
	%\vspace{-0.43cm}	
	\caption{Experimental Testbed}
	\label{Figure:Experimental_Testbed}	
	%\vspace{-0.53cm}	
\end{figure}

The experimental testbed is presented in Figure \ref{Figure:Experimental_Testbed}. The analytics service at each level was implemented using the OpenIoT platform. The OpenIoT components presented in Section \ref{openiot-arch} have been implemented using Java J2EE framework using the Virtuoso RDF triplestore\cite{thakker2010pragmatic}. For more details on the implementation of OpenIoT refer to \textit{www.openiot.eu}.

The OpenIoT system was deployed on two instances of Microsoft Azure servers and one instance of a Google Cloud Server. Table \ref{cloud-spec:table} provides a summary of the server configurations. To test the performance of the system under load, we used Apache JMeter \footnote {http://jmeter.apache.org/} to generate user queries. The OpenIoT instance on windows azure are connected to the sensor platforms producing the data. For experimental purposes, we used a test dataset collected from publicly available weather and pollution data from the year 2014. The total amount of data in the virutoso triple store is around 10 million triples. 

%\begin{figure}[h!]
% \centering
%% \vspace{-0.23cm}
% \includegraphics[scale=0.20]{Testbed.pdf}
%%\vspace{-0.43cm}	
% \caption{Experimentation Testbed}
% \label{Figure:Architecture}	
%%\vspace{-0.53cm}	
%\end{figure}
%

\subsection{Experiment Description}

To evaluate the performance of the proposed hierarchical data analytics system  using  the implemented OpenIoT system on multi-cloud environments, we conduct two experiments.  The OpenIoT instance on the Google Cloud (OpenIoT-1-Google) fetches data from the 2 OpenIoT instances on Windows Azure cloud. The OpenIoT-1-Google server fuses data from the two Azure instances to provide a combined analysis of the data to the end-user. To measure the performance of the system, we use CLAMS \cite{khalid}, a multi-cloud multi-layer performance monitoring framework. CLAMS enables a deep understanding of the performance of each individual component of our hierarchical data analytics systems deployed across the cloud layers e.g. IaaS and PaaS. CLAMS addresses the gaps in existing cloud monitoring tools inability to monitor application deployed in multi-cloud provider environments.

\begin{figure*}[t!]
	\centering
	\begin{subfigure}{0.5\textwidth}
		\centering
		\includegraphics[scale=0.45]{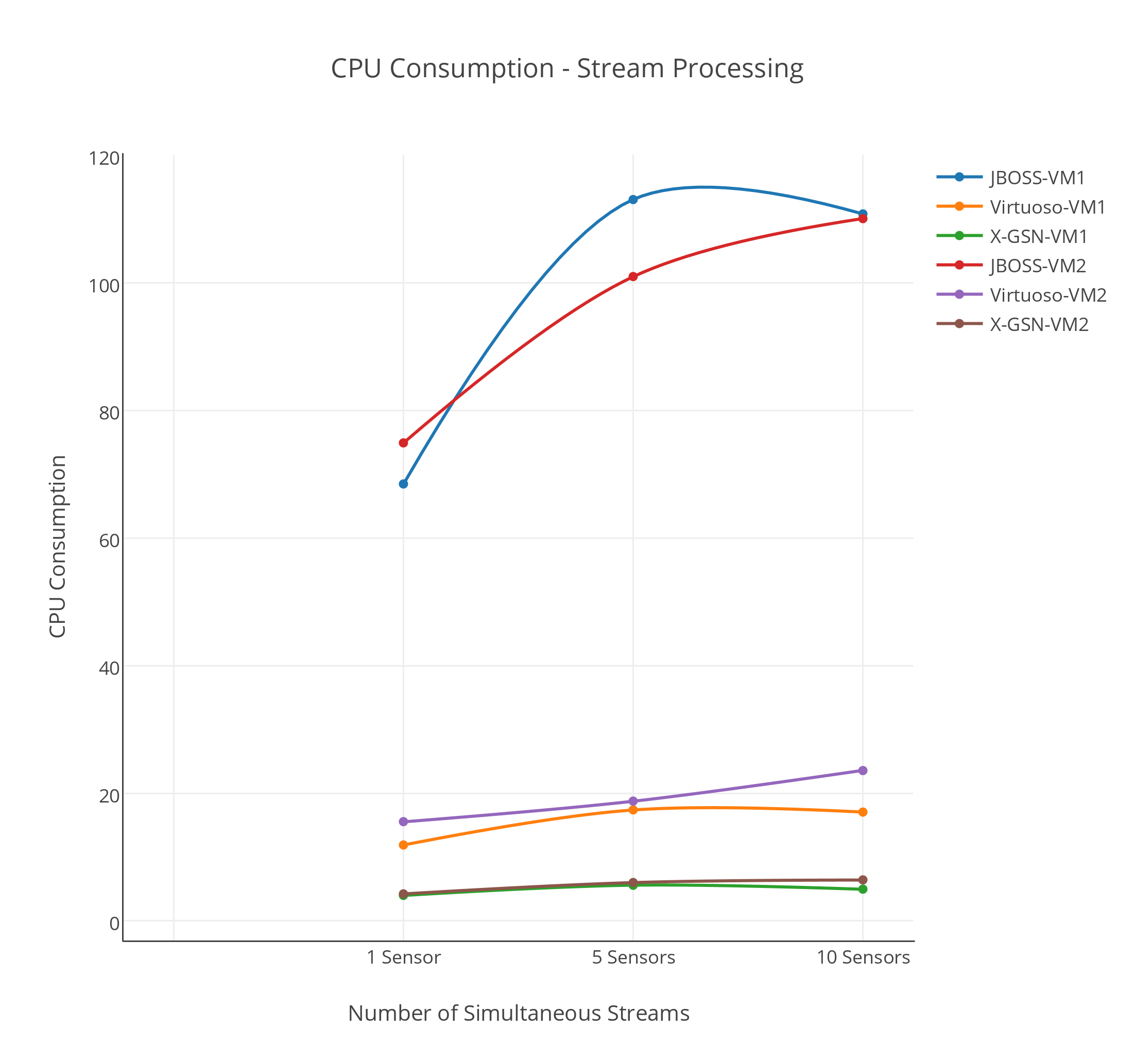}
		\caption{1a: CPU Consumption}
		\label{fig:scpu}
	\end{subfigure}%
	\begin{subfigure}{0.5\textwidth}
		\centering
		\includegraphics[scale=0.45]{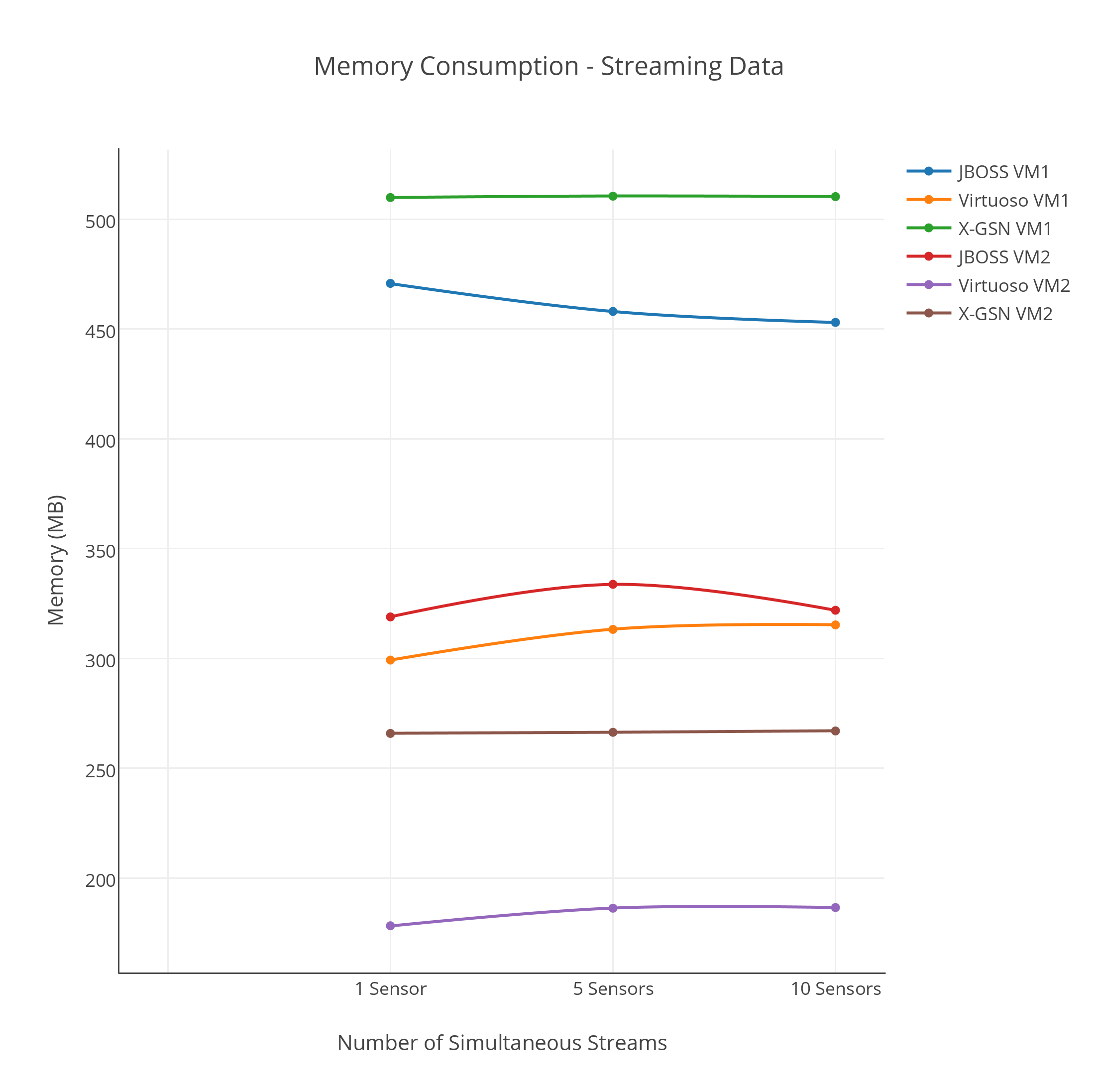}
		\caption{1b: Memory Consumption}
		\label{fig:smem}
	\end{subfigure}
	\caption{Streaming Data Performance}
	\label{fig:streamdata}
\end{figure*}

\textit{Experiment 1 - Streaming Data:} A key to the realisation of the multi-cloud hierarchical data analytics model is its ability to handle streaming data. In this experiment, we use different two cloud configurations namely OpenIoT-1-Azure and OpenIoT-2-Azure. We test the stream data performance by increasing the number of sensors from 1 to 10. Each sensor produces 5 data streams including temperature, humidity, carbon monoxide, pressure and noise. So in total, when 10 sensors are active, the system handles around 50 data streams. The streaming rate is fixed at 1 data point/second. The data generated is time series data i.e. a combination of timestamps associated with data points (double).

\textit{Experiment 2 - Distributed Hierarchical Query Performance:} In this experiment, we measure the response time for query processing. The queries are generated from the Google Cloud OpenIoT instance and are processed distributed by the Azure instances of OpenIoT. 

In both experiments, we also compute the total CPU and memory consumption of each of the OpenIoT component. This provides us with fine grained understanding of the system's performance under load. Each experimental run was repeated 3 times and the results presented here are the average of these outcomes.

\subsection{Experimental Results}
\textit{Experiment 1- Streaming Data Performance:} Figure \ref{fig:streamdata} presents the outcomes of our experiment. The three components that are measured here include JBOSS (hosting all the OpenIoT modules), Virtuoso (the datastore) and X-GSN (the streaming engine connecting sensors to the OpenIoT platform). The results show some interesting observations including CPU consumption of over 100\%. This is due to the fact that in multi-core CPU, when more than one core is used, the CPU consumption goes over 100. For example, in a 4 core CPU, the maximum CPU consumption as reported by CLAMS could be a maximum of 400\%. The VM1 refers to the Azure-1 instance while the VM2 refers to Azure-2 instance. Overall, for managing 50 data streams (10 sensors) at the rate of 1 second, the system performs significantly well without any major bottlenecks. Since the memory consumption of the JBOSS is controlled by the JVM, a trend of higher memory  consumption for VM1 can be noted. This is due to the higher memory availability (7 GB) on VM1 as compared to VM2 (3.5 GB).

\begin{figure*}[t!]
	\centering
	\begin{subfigure}{0.5\textwidth}
		\centering
		\includegraphics[scale=0.43]{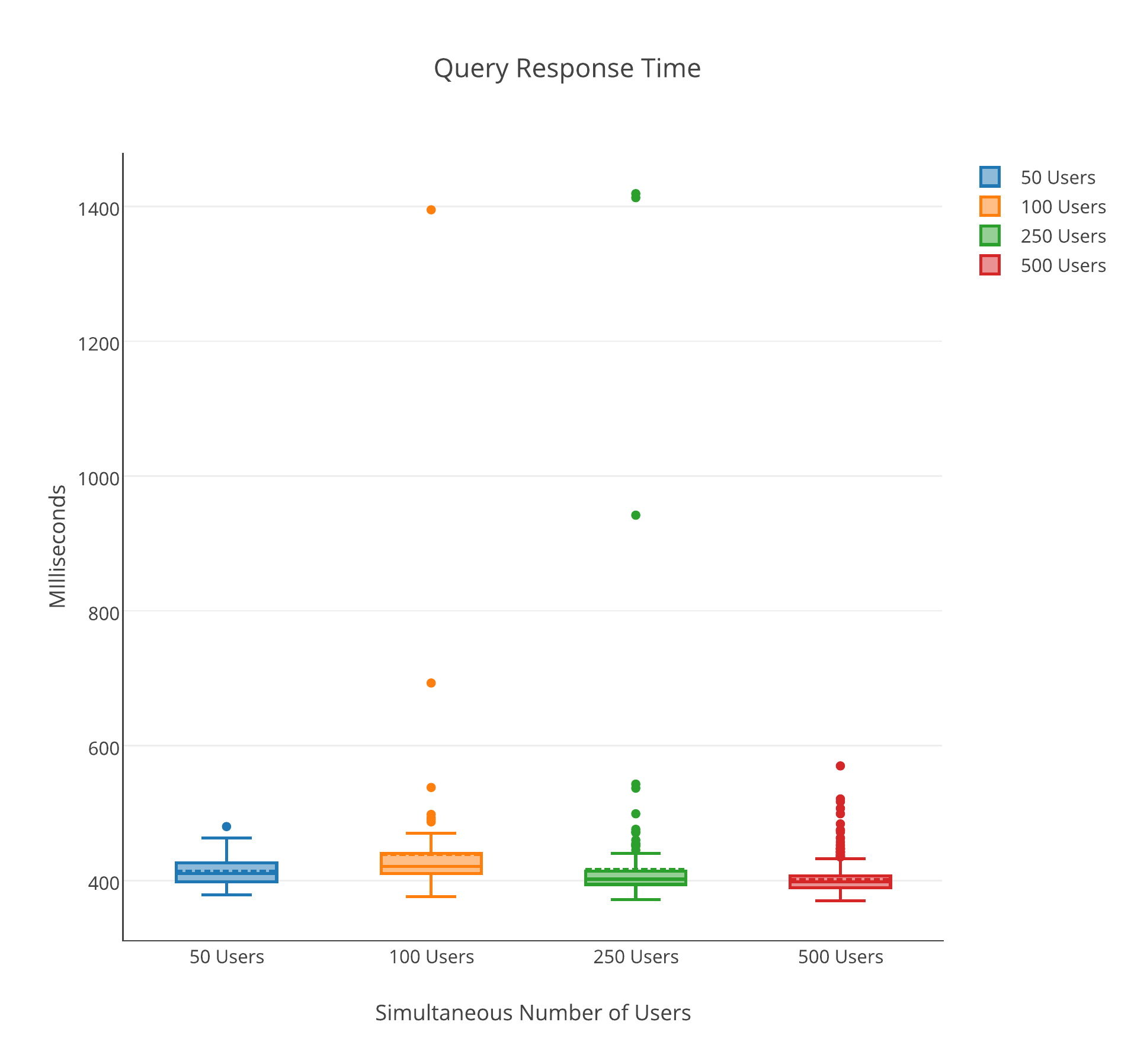}
		\caption{1a: Response Time - OpenIoT-1-Azure}
		\label{fig:res-azure1}
	\end{subfigure}%
	\begin{subfigure}{0.5\textwidth}
		\centering
		\includegraphics[scale=0.43]{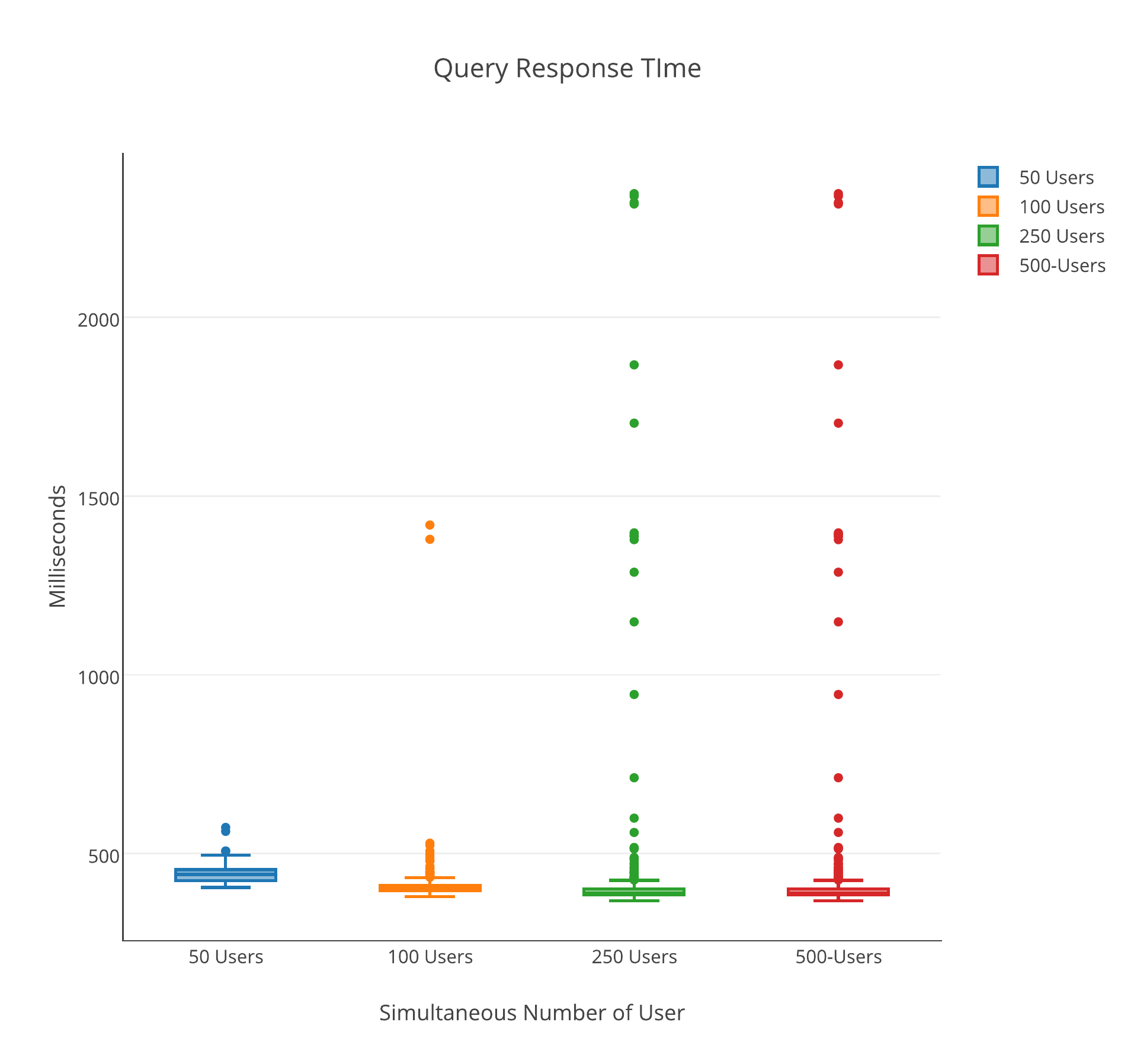}
		\caption{1b: Response Time - OpenIoT-2-Azure}
		\label{fig:res-azure2}
	\end{subfigure}
	\caption{Query Response Times}
	\label{fig:azureresp}
\end{figure*}

\textit{Experiment 2- Distributed Hierarchical Query Performance:} Figure \ref{fig:azureresp} presents the outcome of query response times on the two Azure configuration. The queries originated from the Google Cloud OpenIoT instance. In general, the overall query response time is very good in the order of 400 - 450 millisecond with number of parallel users increasing from 50 to 500. As expected, the Azure 1 instance which has more memory and CPU cores performs better than the Azure 2 instance. The interesting result here is, the response time decreases as number of users increase. This is something we suspect to be associated with how the JVM will allocate memory when the load on the system increases. This outcome is consistent with the outcomes from both the Azure configurations.

\begin{figure*}[t!]
	\centering
	\begin{subfigure}{0.5\textwidth}
		\centering
		\includegraphics[scale=0.43]{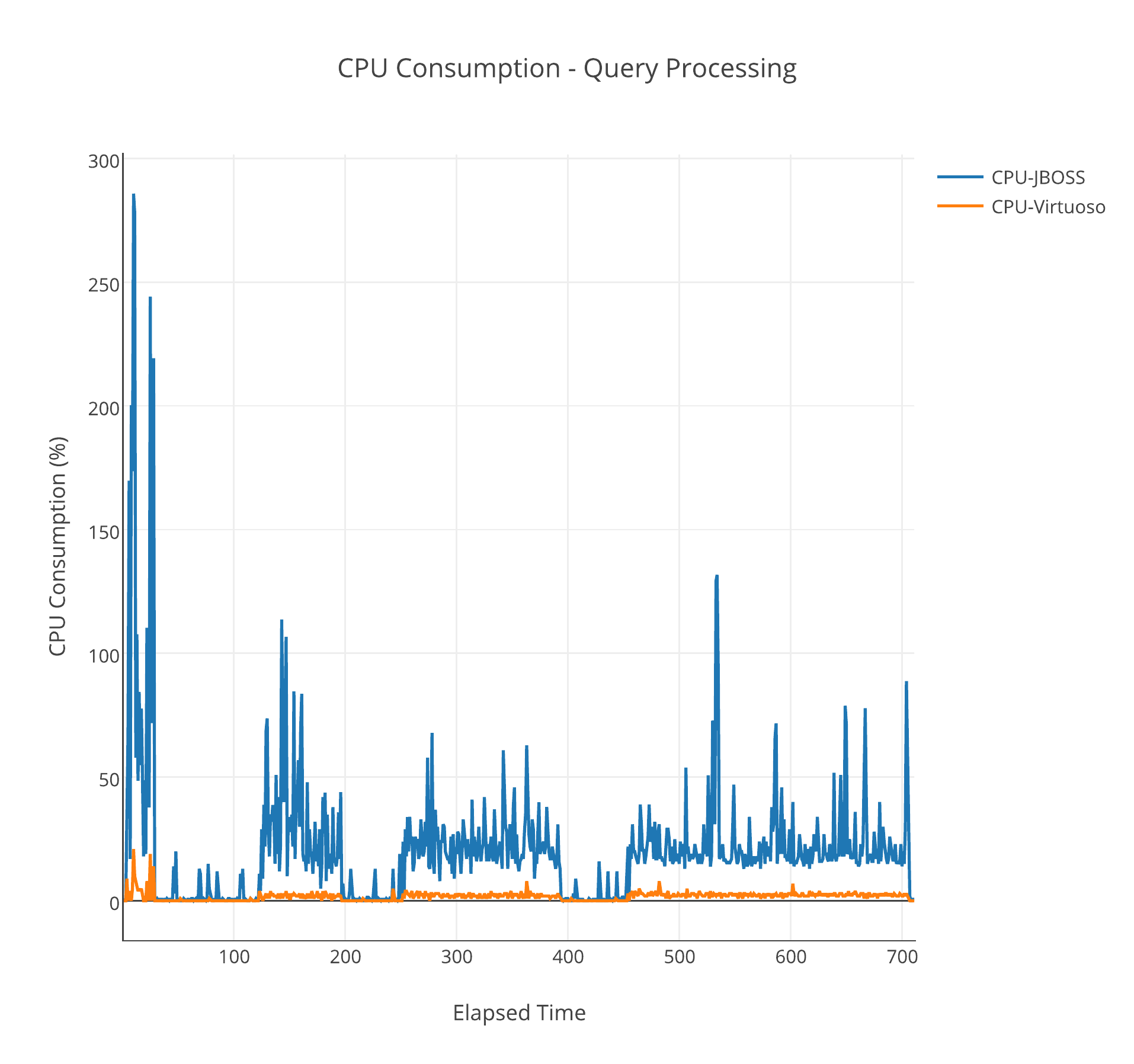}
		\caption{1a: CPU Consumption - OpenIoT-1-Azure}
		\label{fig:cpu-azure1}
	\end{subfigure}%
	\begin{subfigure}{0.5\textwidth}
		\centering
		\includegraphics[scale=0.43]{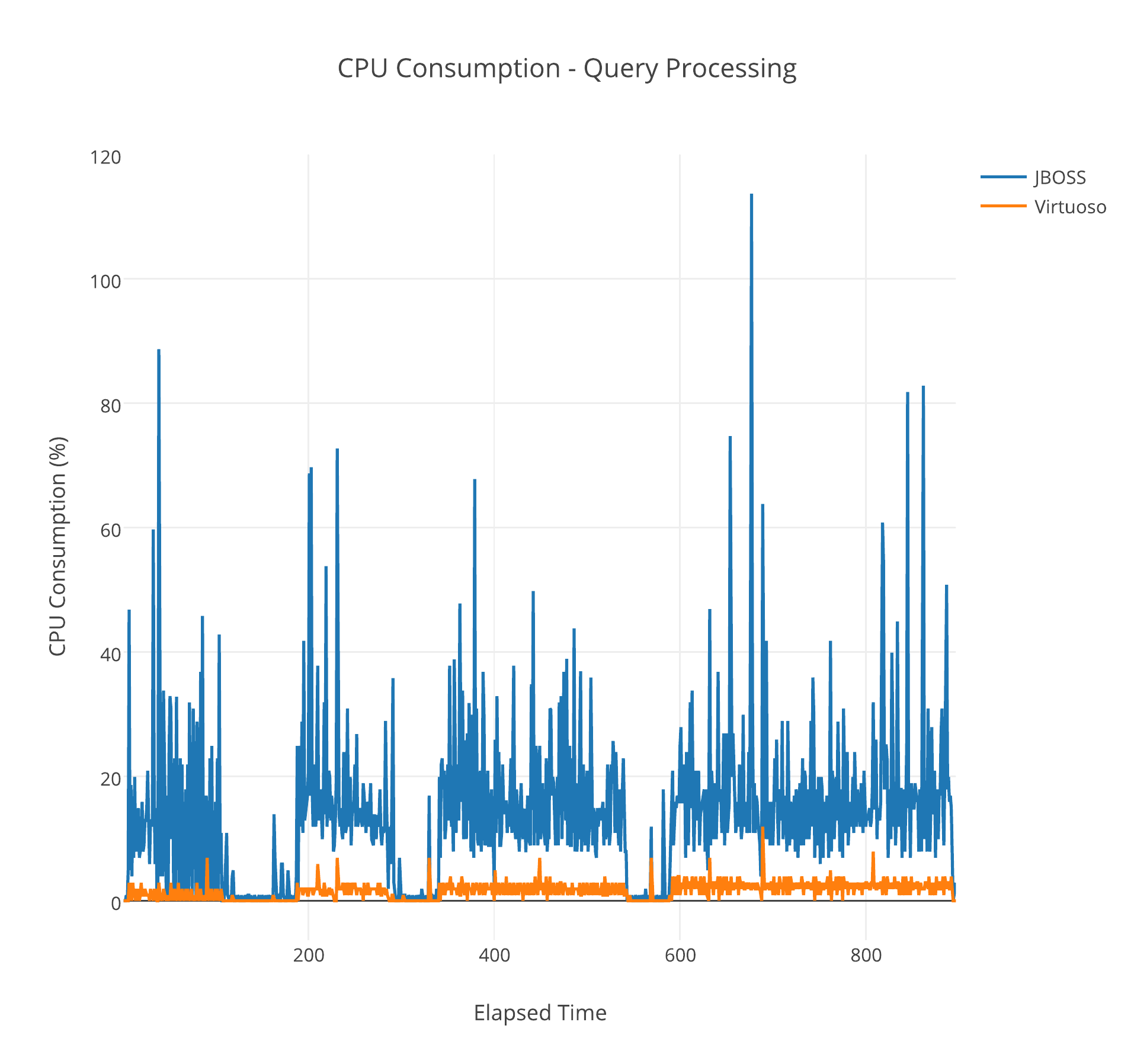}
		\caption{1b: CPU Consumption  - OpenIoT-2-Azure}
		\label{fig:cpu-azure2}
	\end{subfigure}
	
	\begin{subfigure}{0.5\textwidth}
		\centering
		\includegraphics[scale=0.43]{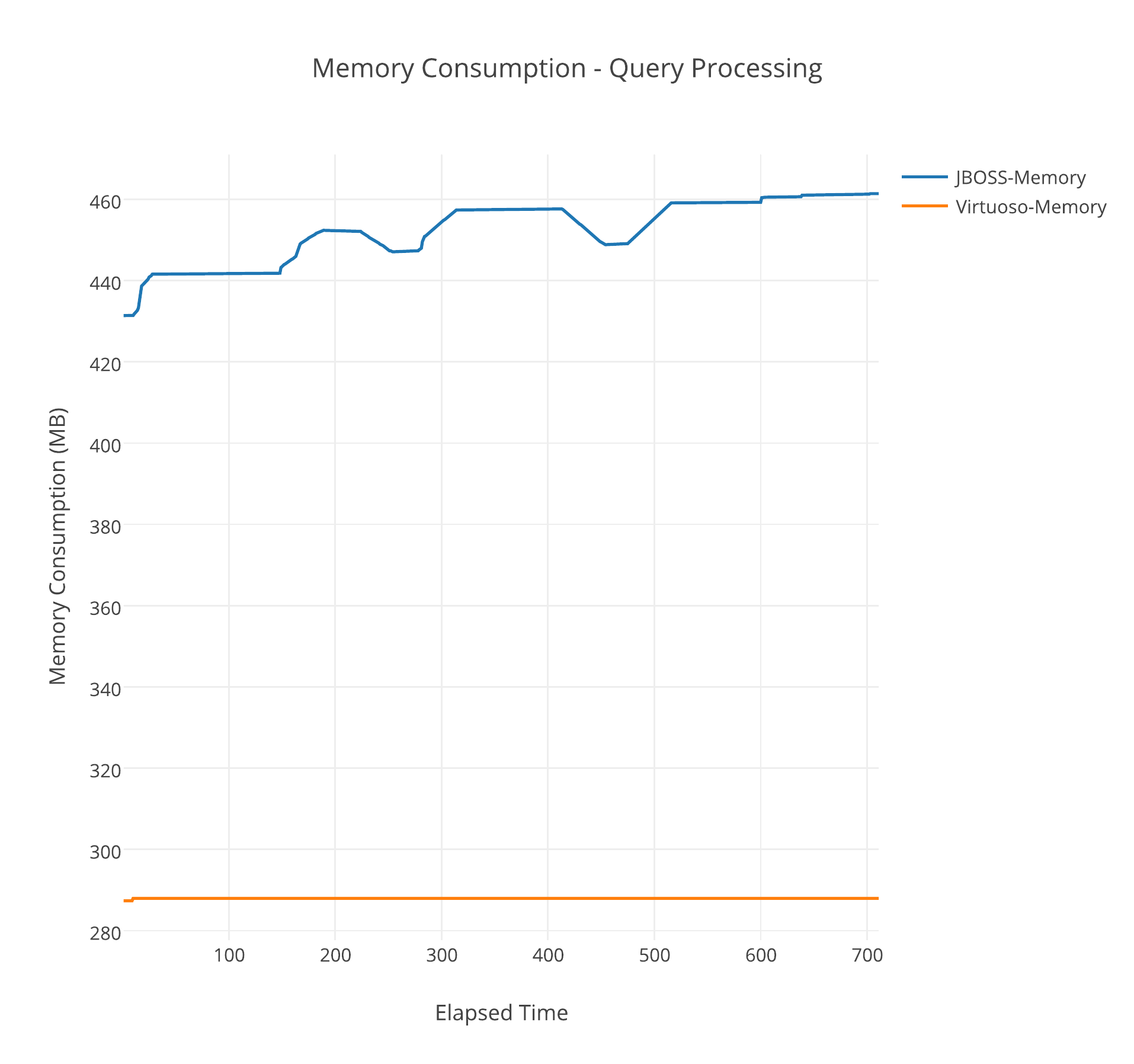}
		\caption{1a: Memory Consumption - OpenIoT-1-Azure}
		\label{fig:mem-azure1}
	\end{subfigure}%
	\begin{subfigure}{0.5\textwidth}
		\centering
		\includegraphics[scale=0.43]{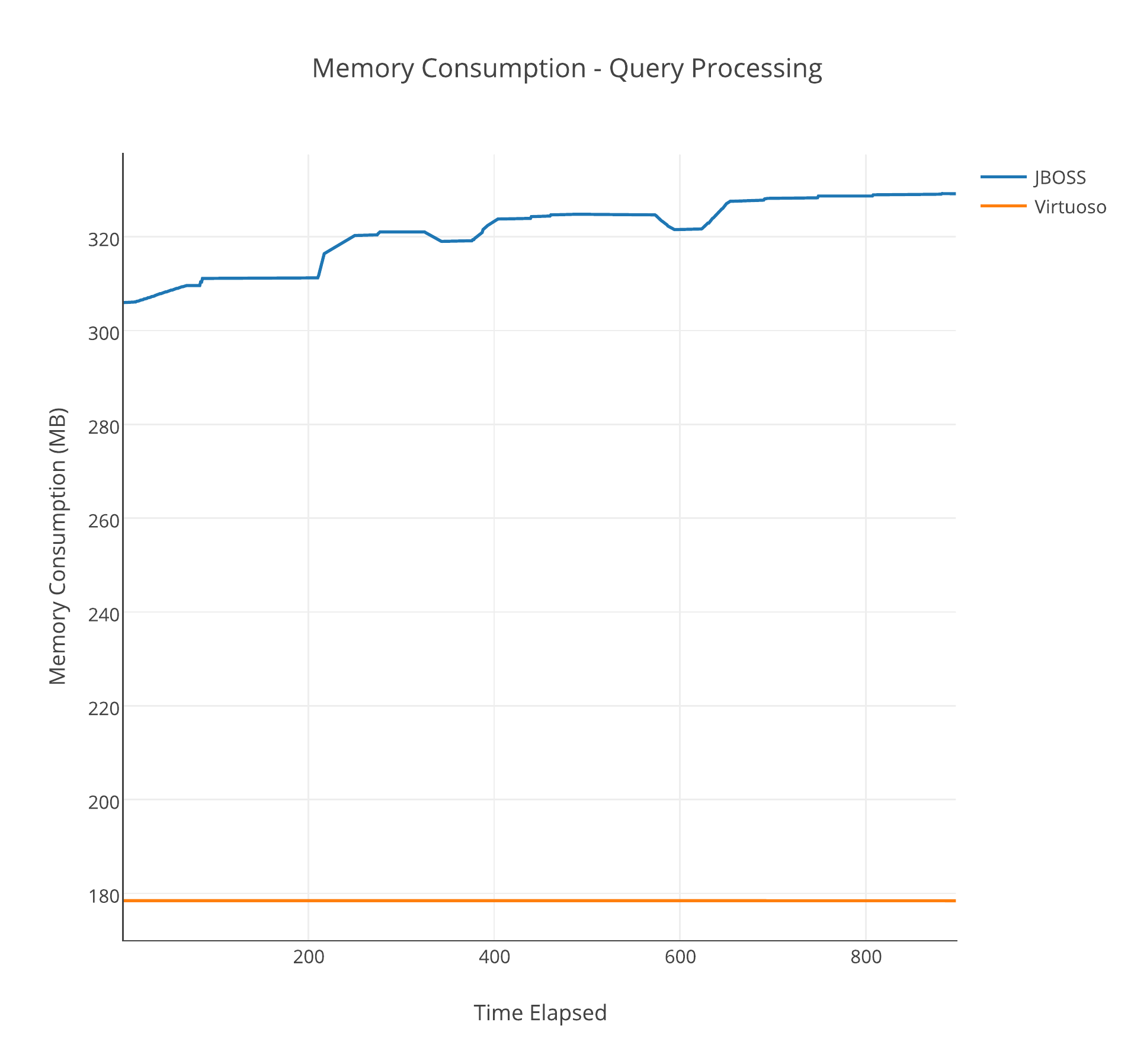}
		\caption{1b: Memory Consumption  - OpenIoT-2-Azure}
		\label{fig:mem-azure2}
	\end{subfigure}

	\caption{Hierarchical Query Processing Performance}
	\label{fig:queryperformace}
\end{figure*}

Figure \ref{fig:queryperformace} presents the CPU and memory consumption of both the Azure 1 and Azure 2 instances while processing the queries from the Google Cloud instance. As described earlier, due to the higher configuration of Azure 1, we note that the JBOSS component of OpenIoT in Azure 1 consumes upto 300\% CPU. The same outcomes is observed with the Memory consumption of JBOSS on each of the instance.

The experimental outcomes validates the following key contributes of the paper namely 1) It is feasible to deploy a hierarchical data analytics system where the various systems could be owned by different providers, 2) Using device and service discovery we can compose multi-cloud data analytics applications, 3) the performance of such a system implemented using the widely used OpenIoT system is scalable and does not show any significant limitations or overheads.

\section{Conclusions and Future Work}

In this paper we have proposed a novel, hierarchical data processing architecture suitable for multi-cloud environments.  This architecture provides flexibility to different parties who host their own cloud IoT platforms to share processed data to reduce computation resource consumption collectively. This also reduces the risks associated in sharing raw data. Such low privacy risks encourage data owners to share their data with third parties where they will use such data for secondary objectives. The demonstrated system is semantically inter-operable. Such interoperability allows different instances deployed in multi-cloud environments to work together to collectively analyse data to achieve a common objective through hierarchical data processing. This was demonstrated in this paper by real-world implementation of the OpenIoT system on Azure and Google cloud platforms. Finally, the experimental results validate the scalability of our proposed multi-cloud data analytics approach. Moreover experimental outcomes also show that the system does not  impose any significant limitations or overheads. Our next step is to develop a complimentary performance model for such hierarchical data processing in multi-cloud environments for autonomous provisioning of cloud resources.

\section*{Acknowledgements}

Charith Perera's work is supported by European Research Council Advanced Grant 291652 (ASAP).

\section*{Competing interests}
The authors declare that they have no competing interests.

\section*{Author's contributions}
Prem Prakash Jayaraman participated in brainstorming, design, experimentation and drafting the manuscript.
Charith Perera and Rajiv Ranjan participated in brainstorming, design and drafting the manuscript. Dimitrios Georgakopoulos, Schahram
Dustdar, and Dhavalkumar Thakker were mentors and contributed to brainstorming, designing and help structure the manuscript.

\bibliographystyle{wileyj} % Style BST file
\bibliography{bmc_article}  
\end{document}